\documentclass[a4paper,10pt,twoside]{cpc-hepnp}

\usepackage{multicol}
\usepackage{graphicx}
\usepackage{booktabs}
\usepackage{amssymb,bm,mathrsfs,bbm,amscd}
\usepackage[tbtags]{amsmath}
\usepackage{lastpage}

\begin{document}

\fancyhead[c]{\small Chinese Physics C~~~Vol. XX, No. X (201X)
XXXXXX} \fancyfoot[C]{\small 010201-\thepage}

\footnotetext[0]{Received 14 March 2009}

\title{Injection performance evaluation for storage ring of SSRF\thanks{Supported by National Natural Science
Foundation of China (11375255) }}

\author{%
      YANG Yong(ÑîÓÂ)$^{1,2;1)}$\email{yangyong@sinap.ac.cn}%
\quad LENG Yong-Bin(ÀäÓñó)$^{1;2)}$\email{lengyongbin@sinap.ac.cn}%
\quad YAN Ying-Bing(ÑÖÓ³±þ)$^{1}$
\quad CHEN Zhi-Chu(³ÂÖ®³õ)$^{1}$
}
\maketitle

\address{%
$^1$ Shanghai Institute of Applied Physics, Chinese Academy of Sciences, Shanghai 201800, China\\
$^2$ Graduate University of Chinese Academy of Sciences£¬Beijing 100049, China\\
}

\begin{abstract}
Injection performance of storage ring is one of the important factors for the light efficiency and quality of Synchrotron Radiation Facility when it is in top-up mode. To evaluate the injection performance of storage ring at SSRF, we build a bunch-by-bunch position measuring system based on oscilloscope IOC. Accurate assessment of energy mismatching, distribution of residual oscillation and angle error of injection kickers can be achieved by this system.
\end{abstract}



\footnotetext[0]{\hspace*{-3mm}\raisebox{0.3ex}{$\scriptstyle\copyright$}2013
Chinese Physical Society and the Institute of High Energy Physics
of the Chinese Academy of Sciences and the Institute
of Modern Physics of the Chinese Academy of Sciences and IOP Publishing Ltd}%

\begin{multicols}{2}

\section{Introduction}

In order to improve the efficiency and quality of the light, Shanghai Synchrotron Radiation Facility adopted top-up mode since the end of 2012, which results in more frequent beam injection(about one injection in ten minutes)~\cite{lab1}. Since injection process involving a variety of equipments, parameters of all the equipments can not achieve perfect matching in the transient process of the injection~\cite{lab2,lab3}. Any parameter mismatching will lead to a closed orbit distortion, which will leave a residual betatron oscillation after injection. For light users, this kind of disturbance must be as little as possible~\cite{lab4,lab5}. An appropriate analysis and diagnosis tool is needed to provide basis for optimizing the parameters of related equipments. Typically, equipment parameter mismatch include the following:(1)the excitation current waveform mismatch (such as amplitude and timing) between kickers; (2)The energy mismatch between booster extracted bunch (injected bunch) and stored bunch in storage ring (center energy mismatch or synchronous oscillation phase mismatch); (3)The installation angle error of injection kickers. Using Libera TBT data and SA data as the analysis tool, only the average effect of bunch train disturbance could be observed, which can analyse issue (3) and part of issue (1), but it becomes invalid when analyse issue (2). When the leakage field of kicker is nonuniform, it is very limited for analysing and optimizing issue (1). This paper comprehensively analyses and studies the problems above and guides equipment parameter optimization by using bunch-by-bunch position measuring system based on oscilloscope IOC for obtaining disturbed details of each bunch.

\section{Brief introduction to injection system of SSRF}

 The layout of injection system of SSRF is shown in Fig.~\ref{fig1}. Four kickers make the orbit bump and the injection point is designed at the bump orbit~\cite{lab6}.
\begin{center}
\includegraphics[width=8cm]{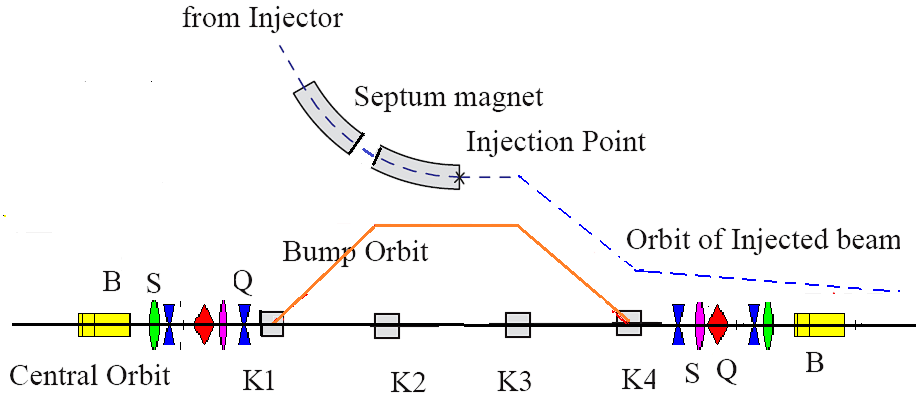}
\figcaption{\label{fig1}   Layout of injection system of SSRF. }
\end{center}

 The excitation current waveform of injection kicker and its timing relation to the bunch train in the storage ring are shown in Fig.~\ref{fig2}(a) and Fig.~\ref{fig2}(c). In the ideal condition, the timing between excitation current waveform and injected bunch must be locked strictly. When the excitation current of 4 kickers match perfectly, which forms an ideal closed orbit with no closed orbit distortion, the injection process is completely transparent for the stored beam and no residual oscillation occurs. However, the excitation current of 4 kickers will mismatch in the actual equipments, as shown in Fig.~\ref{fig2}(b) and residual oscillation occurs because of a time-varying closed orbit distortion~\cite{lab7}. When the storage ring is filled with 500 bunches continuously and uniformly, the closed orbit distortion is not the same for each bunch, so the residual oscillation amplitude on each bunch are not the same and may be completely reversed.
\begin{center}
\includegraphics[width=8cm]{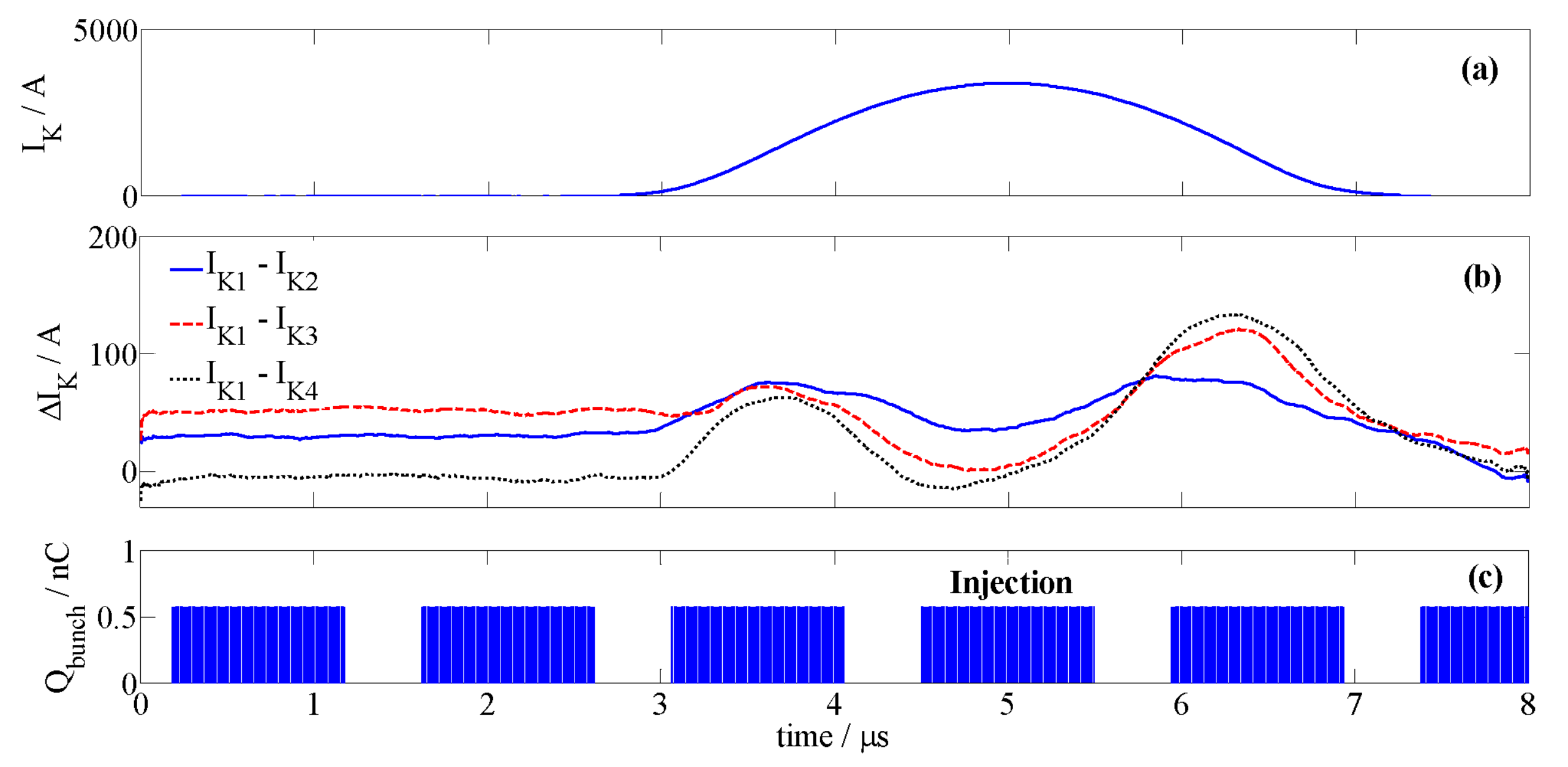}
\figcaption{\label{fig2}   Timing relation between bunch train and excitation current waveform. }
\end{center}

\section{Measurement system setup}

To obtain disturbed details of each bunch, comprehensive analyse and study the problems above and guide equipment parameter optimization,we need to build a bunch by bunch position measuring system.

For SSRF, the bandwidth of this system should be larger than 250MHz, data rates should be greater than 499.654MHz, and the data buffer should be greater than 720*4000 data points (determined by the loop damping time 5-6ms)~\cite{lab8}. Above all these reasons, we use the oscilloscope IOC and  four button electrodes to achieve this system.

Using broadband scope which has high sampling rate (5Gsps), four channels and large memory capacity (102Mpts per channel), we could obtain the raw data from four button electrodes.

To get the position of each bunch which passes through button electrode, the first step is to obtain $T_{rf}$. $T_{rf}$ is the reciprocal of the RF frequency. The most straightforward way to get it is to record the frequency reading of signal generator using for synchronizing the whole accelerator. However, the frequency reading of signal generator is consequentially different from the true frequency. So FFT algorithm is used to obtain RF frequency. To improve the precision, we use zero-padding method which makes the data length 128 times than the raw data. And if the original data length is not the integer powers of two, zero-paddling its length to the integer powers of two and then makes the data length 128 times than it.

The next is to determine the starting point. In consideration of channel time delay mismatch in oscilloscope and the measurement accuracy, we choose the first peak point as the starting point for each channel.

The bunch-by-bunch BPM signal could be obtain from the raw waveform data with the sampling interval $T_{rf}$. Since the sampling point may not be on the original sampling point from oscilloscope, cubic spline interpolation algorithm is used to get the sampling point data. Fig.~\ref{fig3} gives the relationship between sampling point and the original data point.

The bunch-by-bunch position could be got by using difference over sum method (¦¤/¦²). And the sum of four signal is the relative change of each bunch.
\begin{center}
\includegraphics[width=8cm]{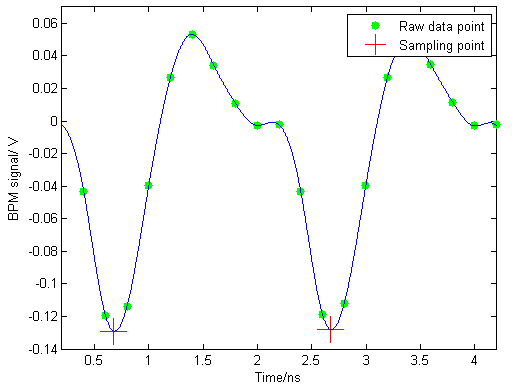}
\figcaption{\label{fig3}   Relationship between sampling point and the original data point. }
\end{center}

\section{Data analysis and process}

\subsection{Data process for sum signal}

The sum signal of four channels, after calibration by DCCT can be used to measure the bunch filling pattern (charge distribution in bunch train), which can evaluate the uniformity of filling pattern~\cite{lab9}.

The refilling bunch index can be identified by the distribution of charge difference before and after injection.

\subsection{Data process for position signal of refilling bunch}

The spectrum of refilling bunch position signal can be obtained by harmonic analysis (red line), as shown in Fig.~\ref{fig4}. The amplitude of resonance peak of energy oscillation can be used to evaluate energy mismatch of refilling bunch. A large amplitude of energy oscillation means a more serious energy mismatch refilled bunch and stored bunch.

As reference, the spectrum of stored bunch is drawn by blue line and the amplitude of energy oscillation is below the noise floor of the measurement system.
\begin{center}
\includegraphics[width=8cm]{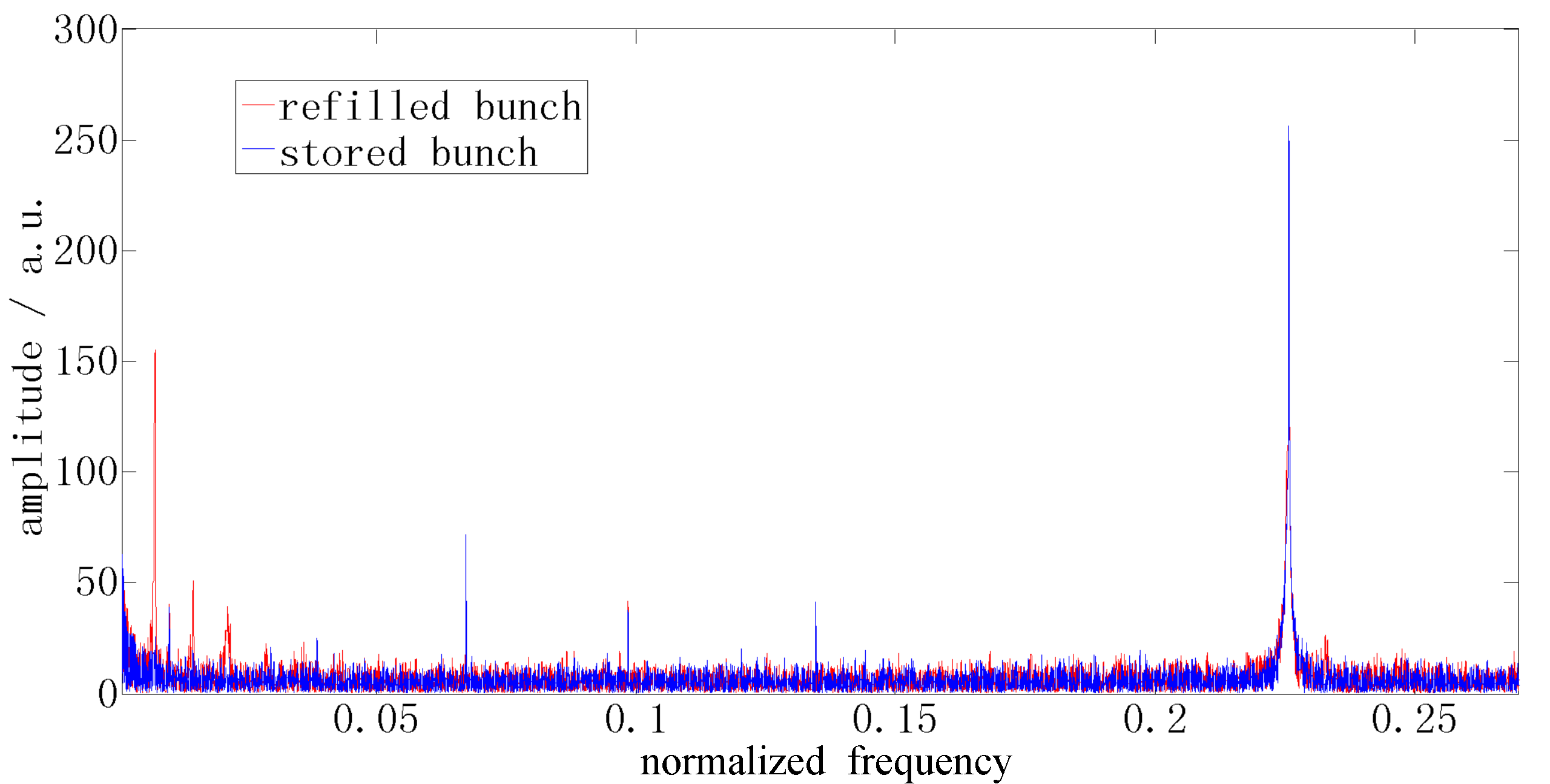}
\figcaption{\label{fig4}   Spectrum of refilling bunch and stored bunch position signal. }
\end{center}

\subsection{Data process for position signal}

The PCA method is used for physical model separation and noise reduction~\cite{lab10}. The space vector of damping betatron oscillation mode (mode 1) corresponds to the residual oscillation amplitude of each bunch (i.e. The mismatch of kickers) as shown in Fig.~\ref{fig5}. Fig.~\ref{fig5} also shows the betatron oscillation mode contributed by wakefield (mode 2) which will not be discussed in this paper.
\begin{center}
\includegraphics[width=8cm]{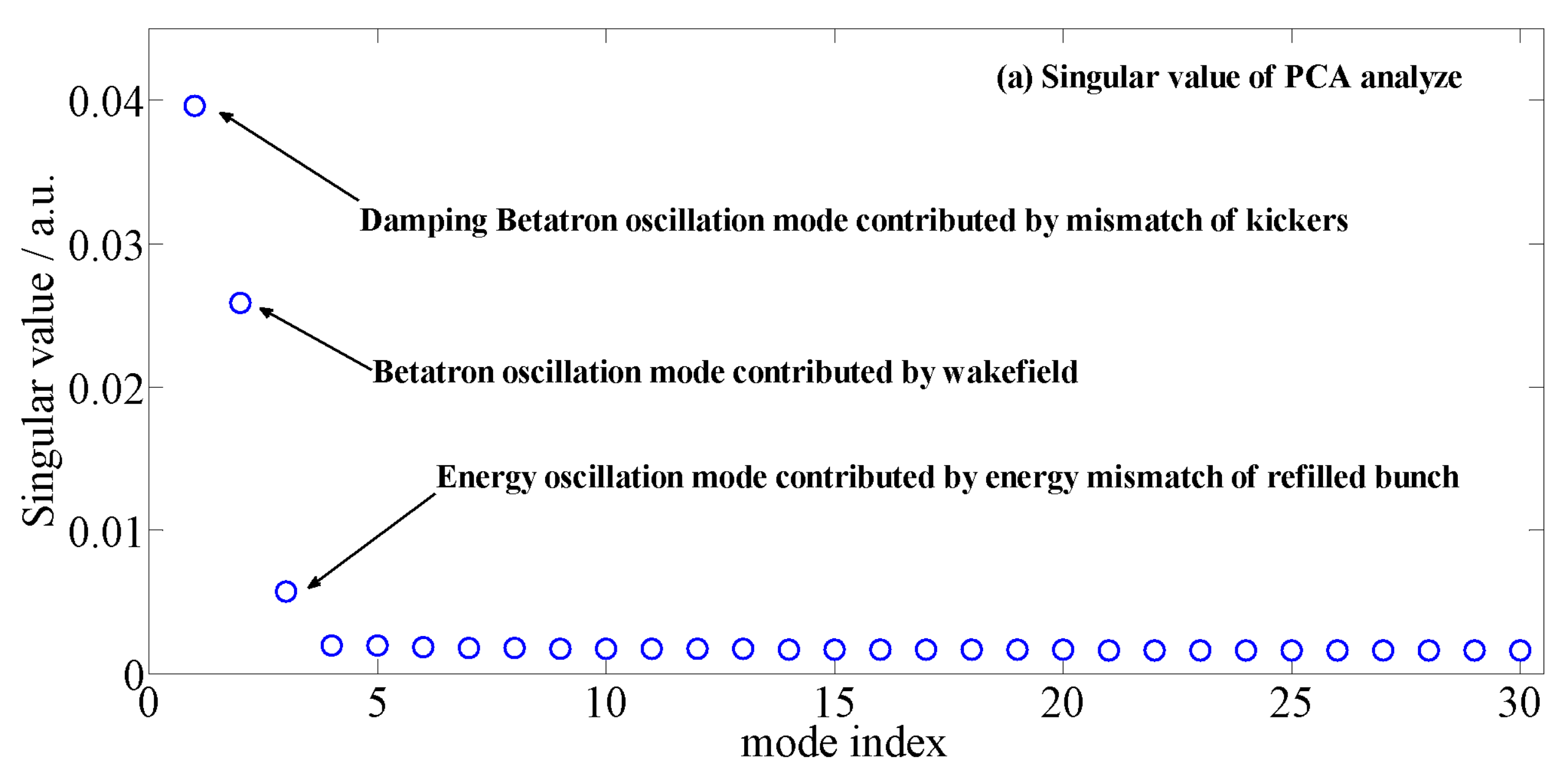} \\
\includegraphics[width=8cm]{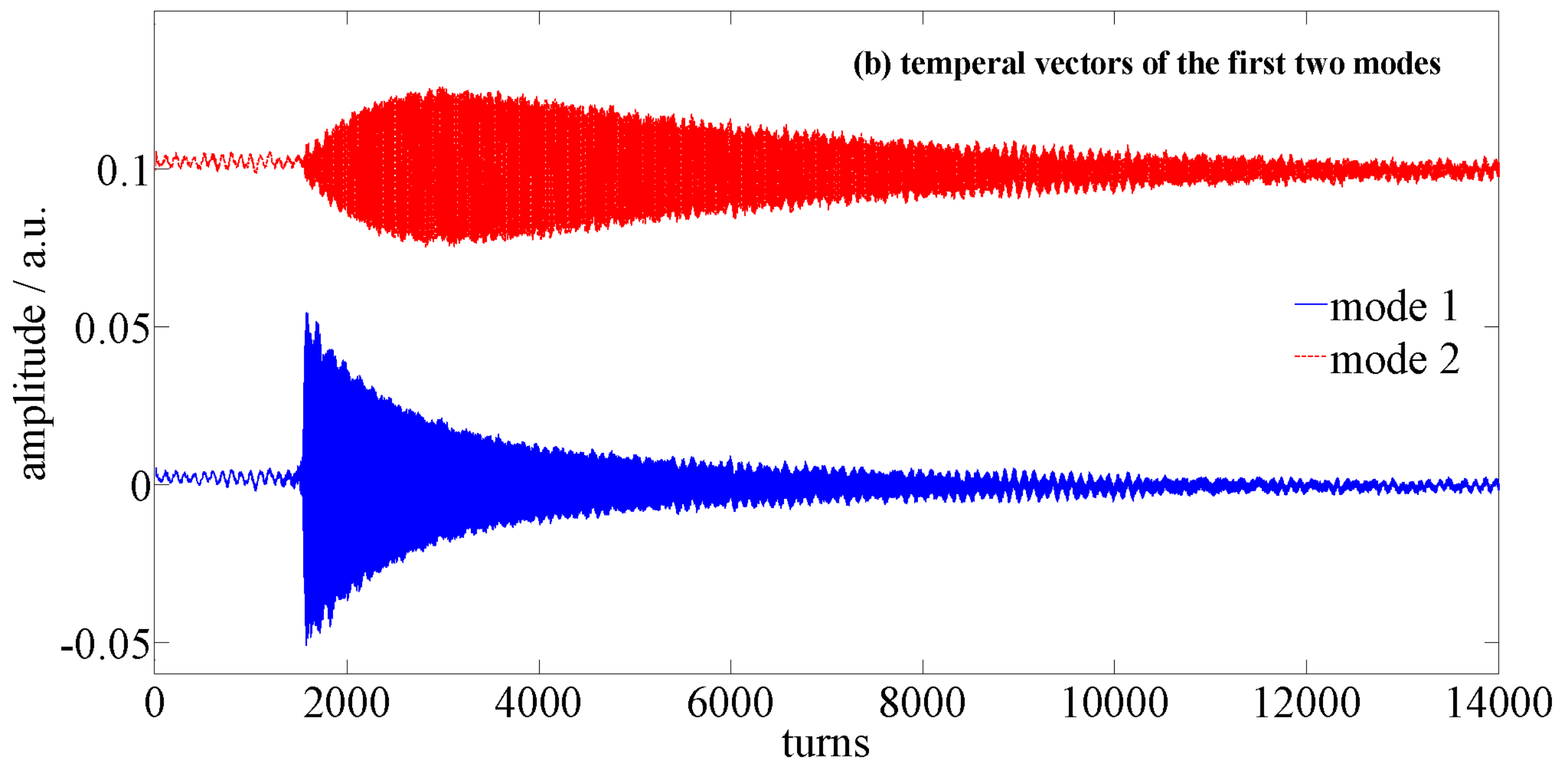}
\figcaption{\label{fig5}   (a) Singular value of PCA analyse for position signal. (b) Temperal vectors of the first two modes. }
\end{center}

\subsection{Reconstruction of amplitude distribution of residual oscillation for all the 720 buckets}

The ideal experimental method is using single-bunch operation and makes the starting point of kicker waveform iterate through all the buckets. However, that will need a large number of machine study time.

Considering that the filling pattern is continuous with 500 bunches when SSRF is in operation at user time, information of 500 buckets can be got during one injection.

Since the timing between excitation current waveform of kicker and injected bunch is strictly synchronous, when the difference of injected bunch index is greater than 220, a complete residual oscillation distribution in one turning period can be obtained by splicing two groups of injection data, as shown in Fig.~\ref{fig6}.
\begin{center}
\includegraphics[width=8cm]{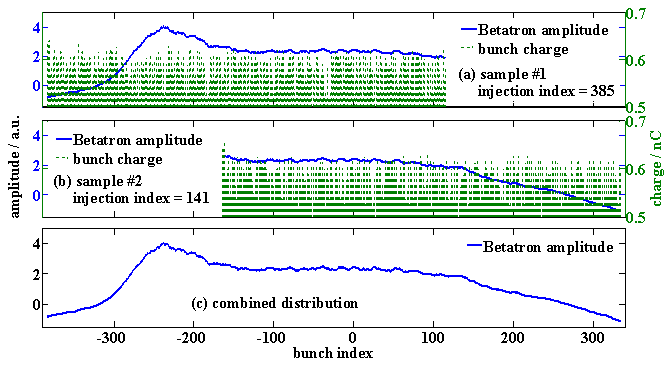}
\figcaption{\label{fig6}   Combine the distribution of residual oscillation. }
\end{center}

\section{Injection performance evaluation for storage ring}

\subsection{Evaluation results for kicker angle and signal delay performance among kickers}

Turn-by-turn data, regarded the bunch train as a whole, which is got by charge weighted average algorithm from bunch-by-bunch data, can be used to observe the residual oscillations in the $\mu s$ scale and evaluate the average performance of the injection system. This measurement is mainly for the optimization and evaluation of kicker angle and signal delay time among kickers.
\begin{center}
\includegraphics[width=8cm]{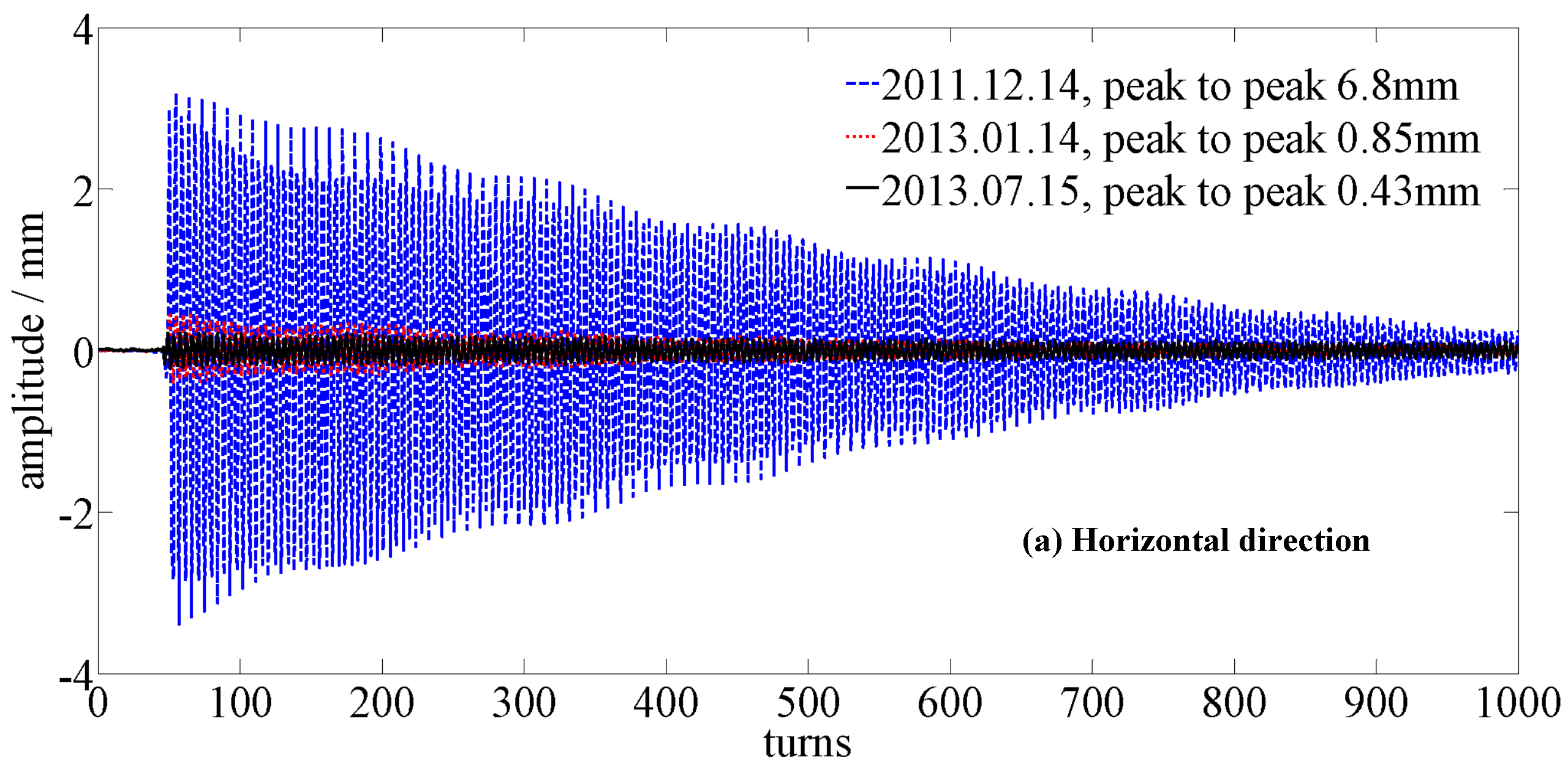} \\
\includegraphics[width=8cm]{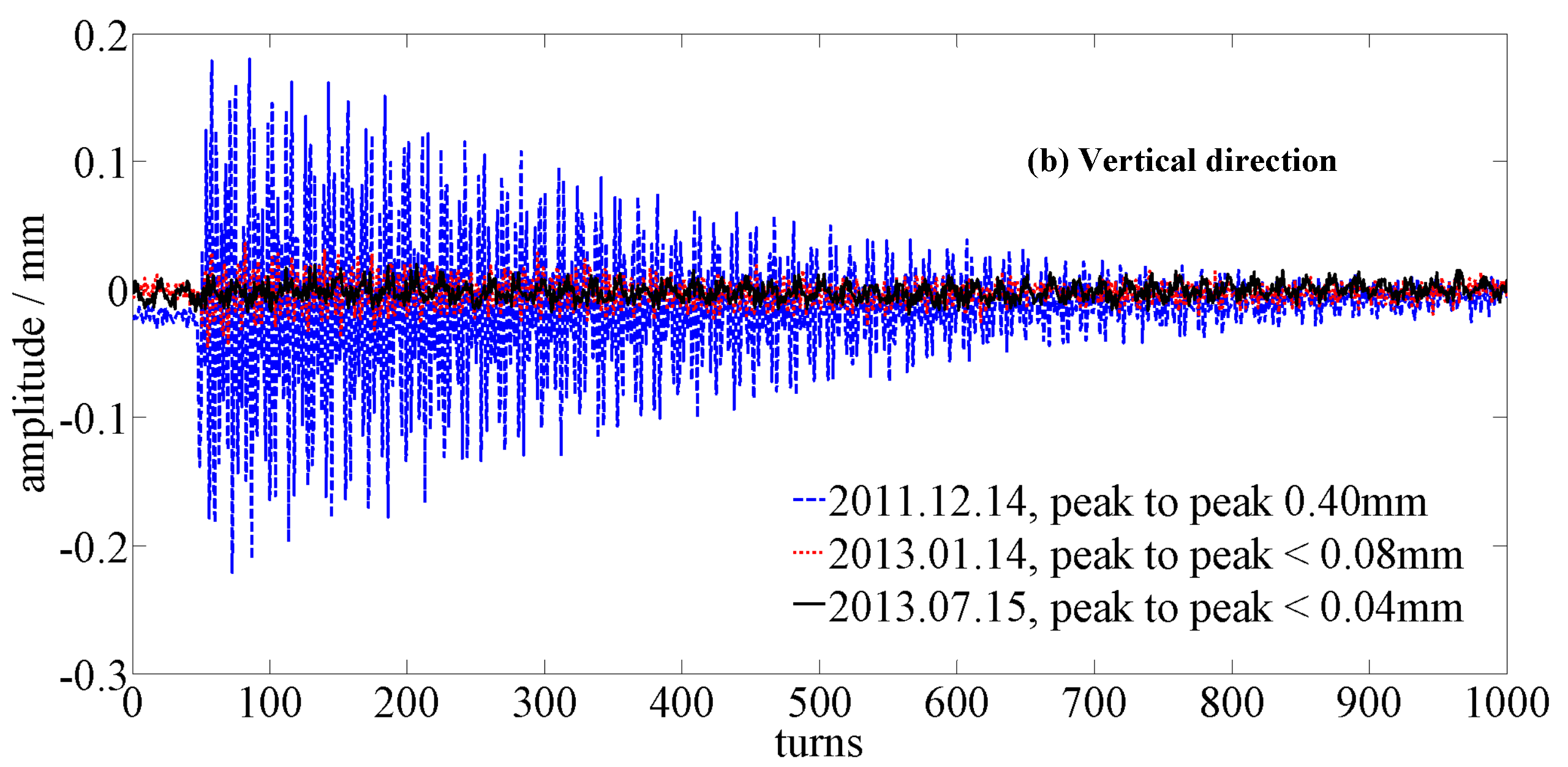}
\figcaption{\label{fig7}   Results of two optimization after 2011. }
\end{center}

Two optimization after 2011 achieved remarkable results to decrease the average residual oscillation amplitude as shown in Fig.~\ref{fig7}.

\subsection{Performance evaluation for energy matching of refilled bunch}
The energy oscillation of refilled bunch is great in SSRF. The oscillation amplitude and refilled charge has a strong linear dependence, which proves the disturbance comes from the energy mismatch between the new charge and the stored charge, as shown in Fig.~\ref{fig8}. This disturbance has no obvious effect to the majority of users, but users who is sensitive to fine structure of the light pulse can observe the disturbance of energy oscillation.
\begin{center}
\includegraphics[width=8cm]{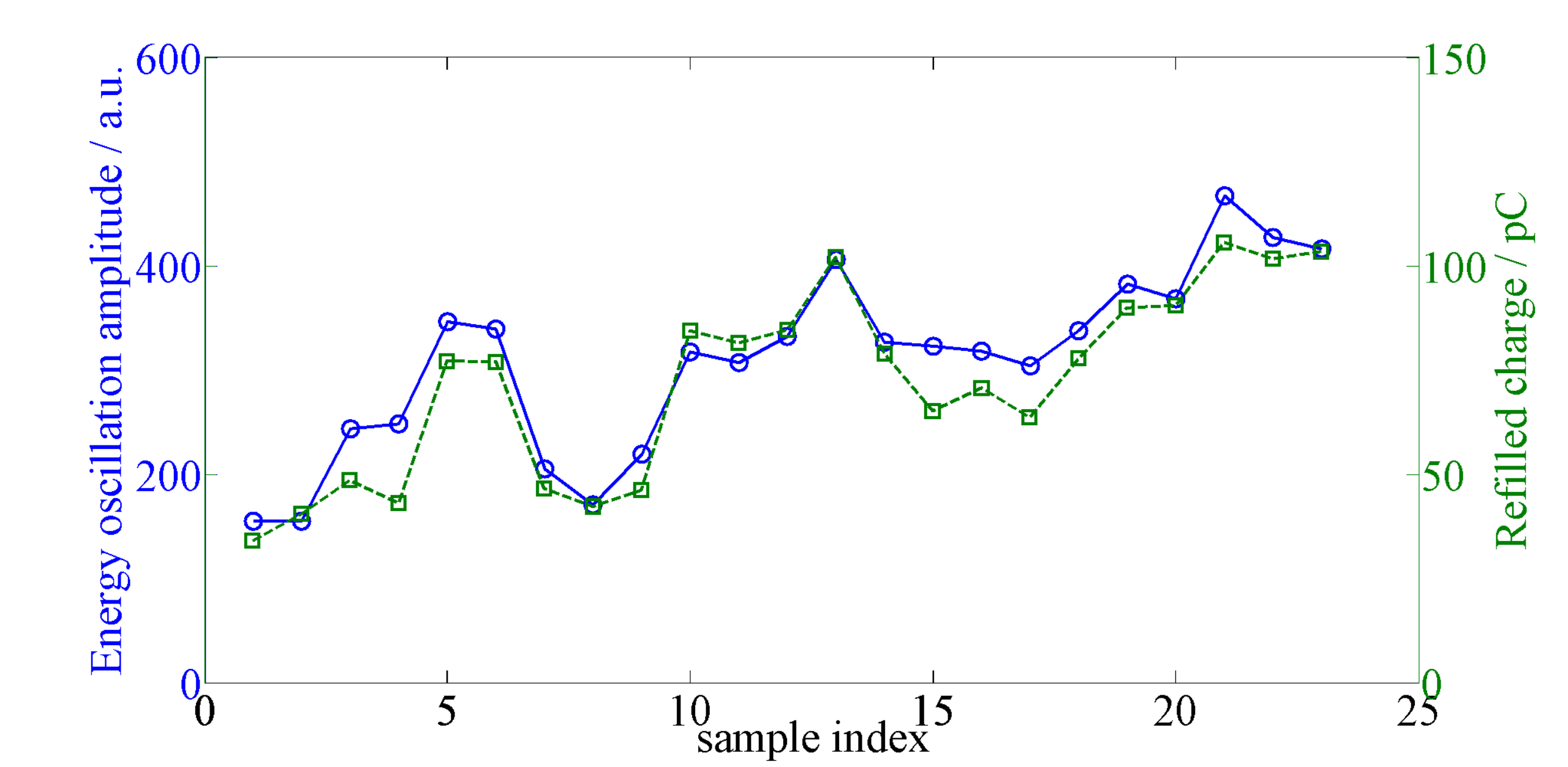}
\figcaption{\label{fig8}   Relationship between oscillation amplitude and refilled charge. }
\end{center}

The fluctuation of energy oscillation amplitude is large between different injections (Fig.~\ref{fig8} shows the amplitude difference can reach more than twice), which is mainly from the fluctuation of refilled bunch charge. This index can be improved by optimizing energy matching degree of the injection bunch and stabilizing injection bunch charge.

\subsection{Determination of the distribution of residual oscillation}

To evaluate the reproducibility, we select three days¡¯ data as shown in Fig.~\ref{fig9}. We can find that the distribution is basically the same, but there is the difference in amplitude. the result means there is random fluctuation to the disturbance of the bunch, which is not conducive to build feed-forward compensation.
\begin{center}
\includegraphics[width=8cm]{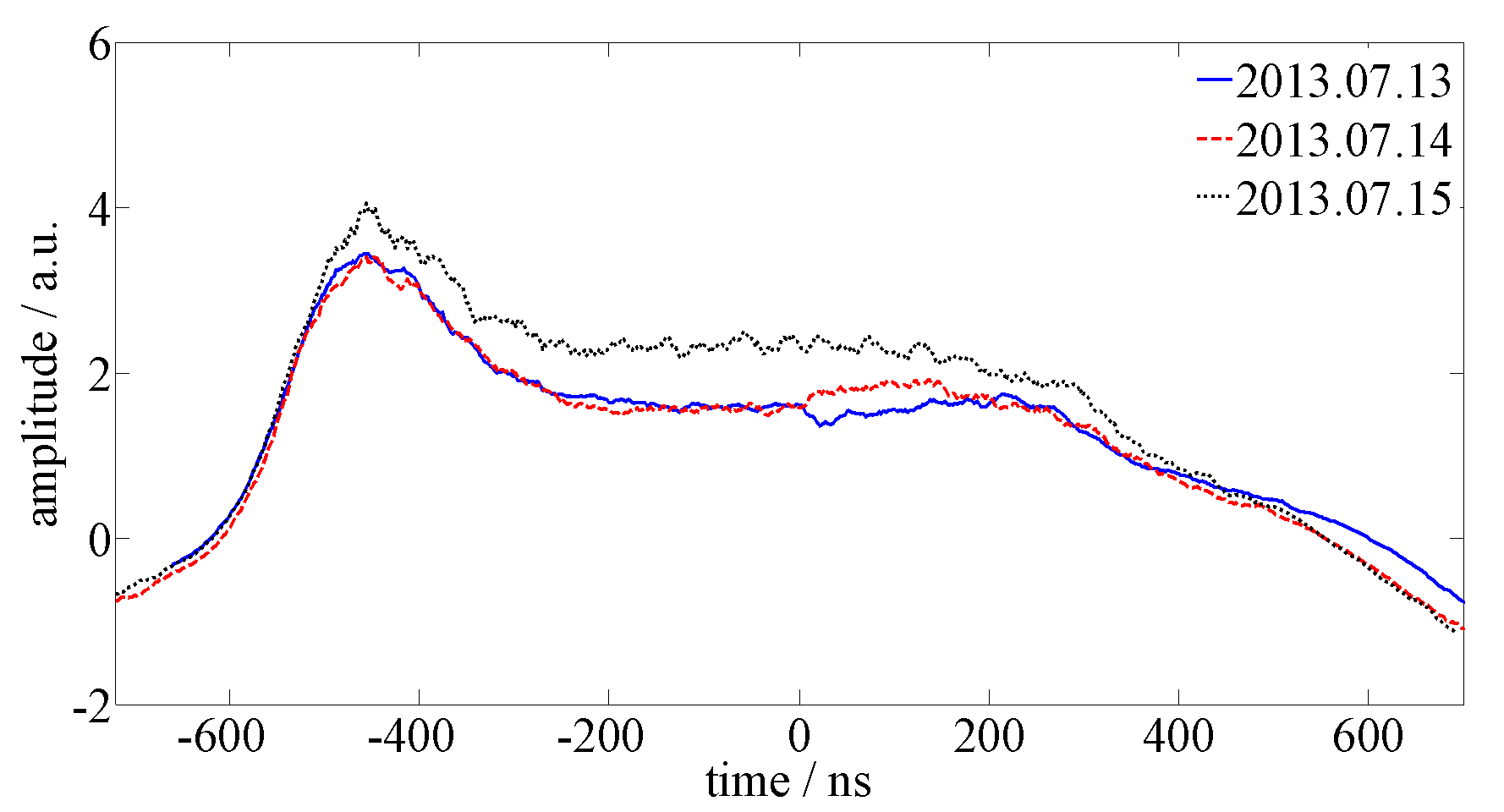}
\figcaption{\label{fig9}   Reproducibility of the distribution of residual oscillation. }
\end{center}

\subsection{Effects on bunch train by non-uniform distribution of residual oscillations}

Regarded the bunch train as a whole, the average residual oscillation amplitude is the charge weighted average of each bunch oscillation amplitude in the storage ring. Based on the distribution of residual amplitude above, the relationship between average residual amplitude and injection bunch index is shown in Fig.~\ref{fig10} when storage ring is continuously filled with 50, 100, 200 and 500 bunch.
\begin{center}
\includegraphics[width=8cm]{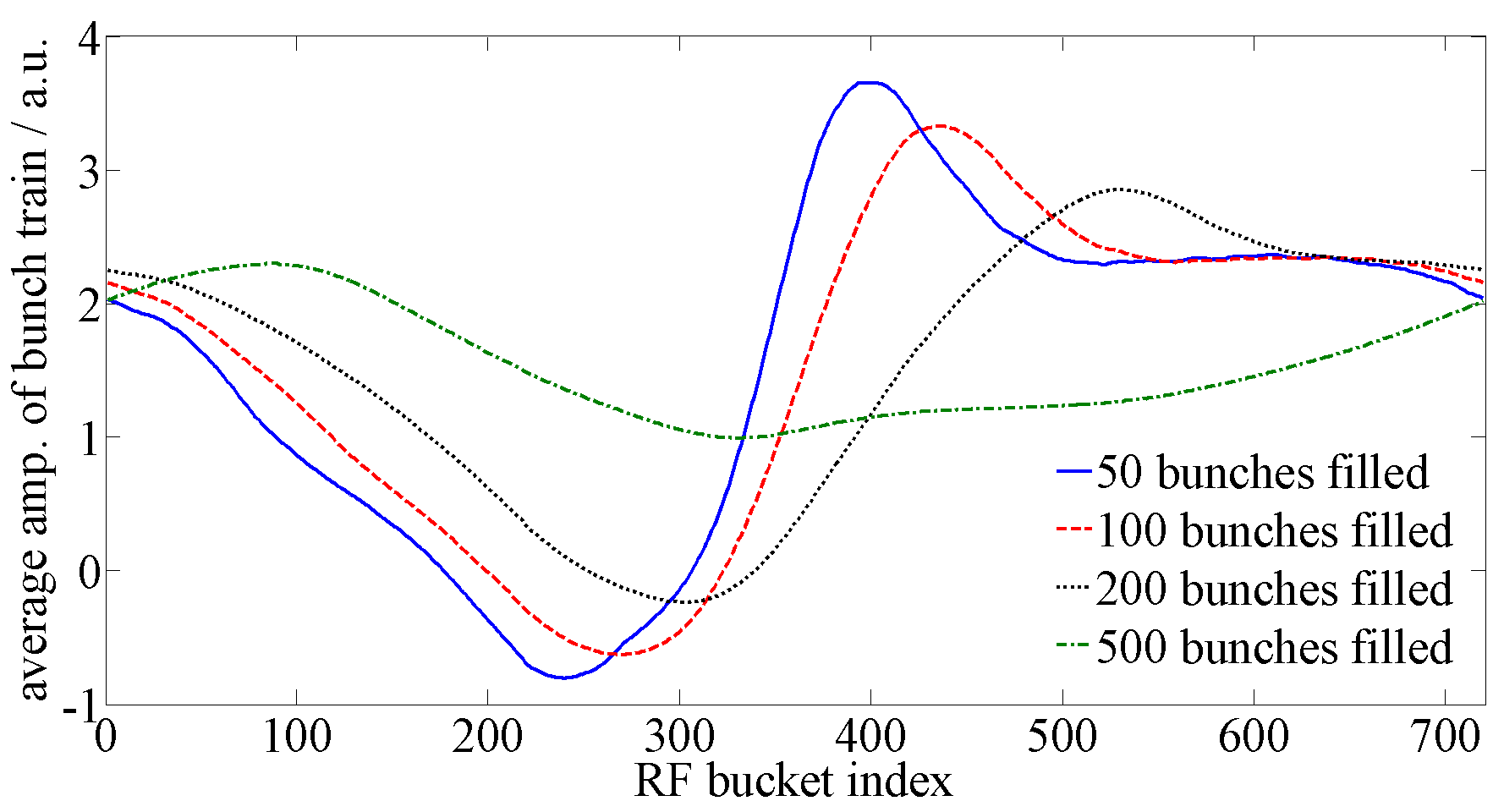}
\figcaption{\label{fig10}   Relationship between average residual amplitude and injection bunch index. }
\end{center}

In Fig.~\ref{fig10}, we can find that the average residual amplitude has a periodic structure, i.e. the disturbance to the bunch train is not a constant value during injection and that is not favorable to the feed-forward compensation. This inference should be observed in the process of beam accumulation.
\begin{center}
\includegraphics[width=8cm]{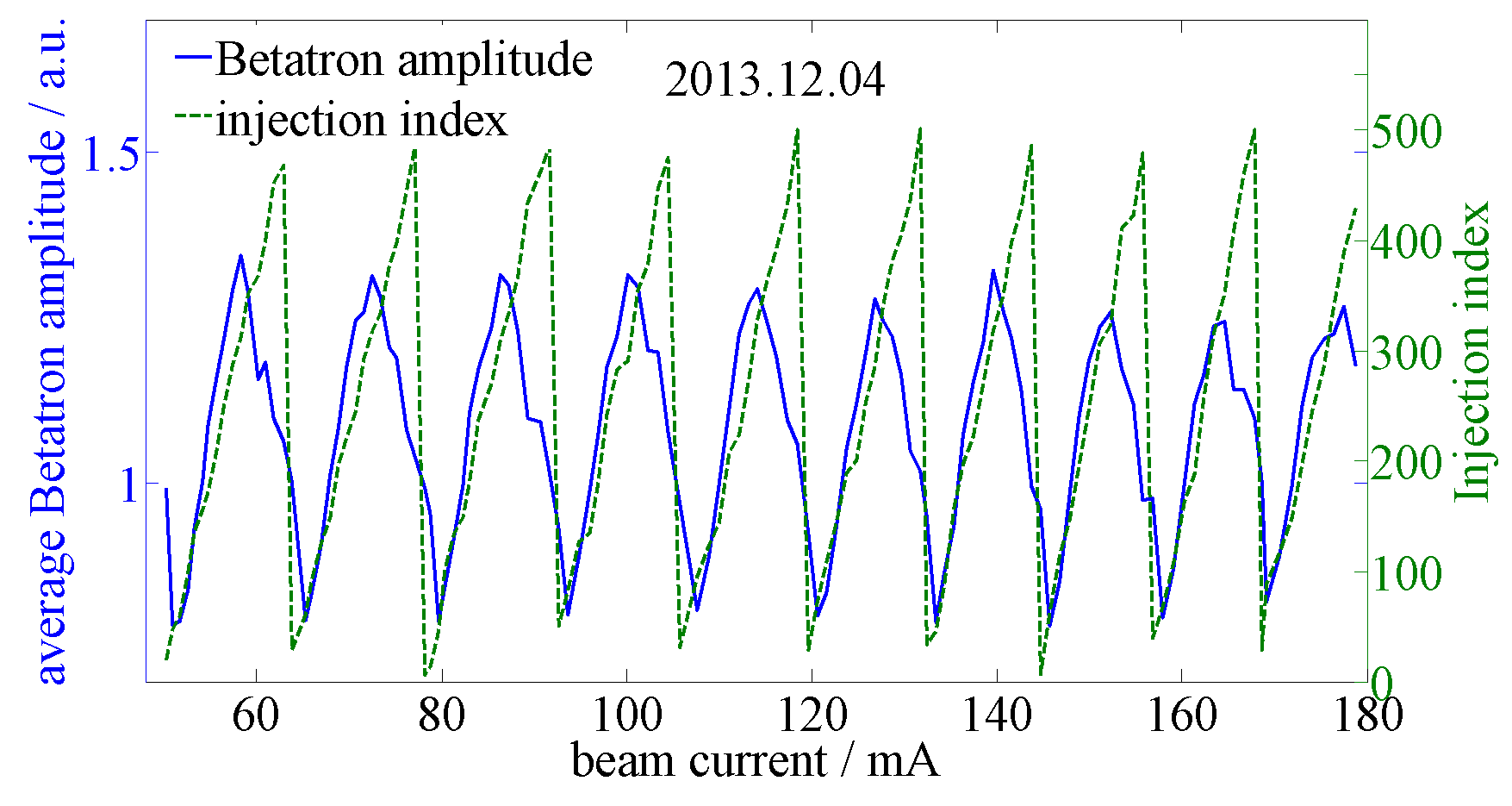}
\figcaption{\label{fig11}   Relationship among betatron oscillation amplitude, injection bunch index and average current. }
\end{center}

To verify the inference, we use Libera to get turn by turn data in the process of beam accumulation. In December 4, 2013, the beam current in the storage ring of SSRF was accumulated from 50mA up to 179mA (continuous filled with 500 bunches). During this process, we recorded the average current, charge distribution of bunch train and turn-by-turn transverse position data at the same time. After data process, the relationship among betatron oscillation amplitude, injection bunch index and average current is shown in Fig.~\ref{fig11}. We can see that there is obvious dependency relationship between the average amplitude of betatron oscillation and the injection bunch index.

Rearranging the data above with taking injection bunch index from low (1) to high (500) as one cycle, the whole process of the beam accumulation can be divided into 10 cycles. Using injection bunch index as the abscissa axis and the amplitude of average residual oscillation as ordinate, we can get the dependency relationship between the two, as show in Fig.~\ref{fig12}. It shows that the average residual oscillation amplitude is a periodic function of injection bunch index. The reproducibility is very good among the 10 cycles, and it is consistent with expectations.
\begin{center}
\includegraphics[width=8cm]{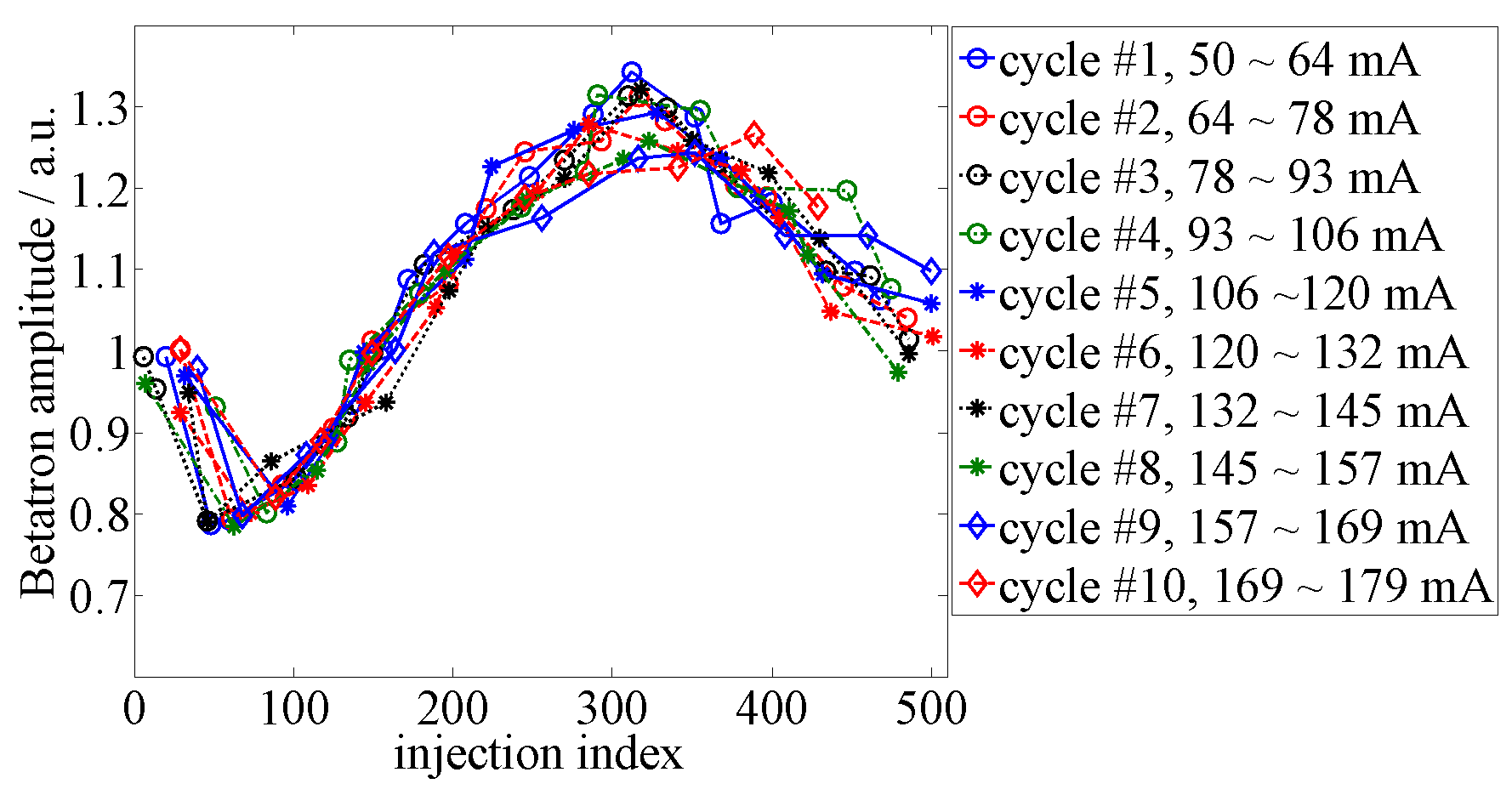}
\figcaption{\label{fig12}   Relationship between amplitude of average residual oscillation and injection bunch index in 10 cycles. }
\end{center}

\section{Conclusion}
Bunch-by-bunch position measurement system as analysis and diagnosis tool is very effective for performance evaluation of injection process in SSRF. Compared with turn-by-turn data, it can obtain much more transient information of injection process and is more conducive to optimize equipment.

We achieve the evaluation of energy matching degree of the injection bunch. The result shows that the energy matching degree of the refilled bunch is poor. The fluctuation of refilled bunch charge is large. The injector still has some optimization space.

Reconstruction of amplitude distribution of residual oscillation for the whole turning period caused by kicker leakage field is accomplished. According to the measurement, the distribution of this amplitude has a periodic structure. The average disturbing amplitude to the stored bunch train depends on the injected bunch index, and it also has a periodic structure. This inference is supported by turn-by-turn experimental data, which agree very well.

The analysis of multiple sets of data indicates the shape distribution of kicker leakage field is stable, but there exists slow drift in amplitude, which is not conducive to the feed-forward compensation. More efforts are also needed to improve the stability of excitation current waveform of kicker.

\end{multicols}

\vspace{-1mm}
\centerline{\rule{80mm}{0.1pt}}
\vspace{2mm}

\begin{multicols}{2}

\end{multicols}

\clearpage

\end{document}